\documentclass{llncs}

\usepackage[T1]{fontenc}
\usepackage[utf8x]{inputenx}

\usepackage{cite}
\usepackage[dvipsnames]{xcolor}
\usepackage{graphicx}
\usepackage{xspace}
\usepackage[hyphens]{url}
\usepackage[breaklinks=true]{hyperref}

\usepackage{array}
\usepackage{enumitem}
\usepackage{upgreek}
\usepackage{multirow}
\usepackage{tabularx}
\usepackage{amssymb}
\usepackage{amsmath}
\usepackage{cleveref}
\usepackage{booktabs}
\usepackage{xifthen}

\usepackage{floatrow}
\newfloatcommand{capbtabbox}{table}[][\FBwidth]

\usepackage{tikz}
\usetikzlibrary{calc,automata,positioning,decorations.pathreplacing}
\tikzset{align at top/.style={baseline=(current bounding box.north)}}
\tikzstyle{every node}=[font=\scriptsize]
\tikzstyle{state} = [draw,fill=white,circle,thick,align=center,inner sep=0pt,minimum size=4.5mm]
\tikzstyle{lstate} = [draw,fill=white,rectangle,rounded corners,thick,align=center,inner sep=2pt]
\tikzstyle{dot} = [fill,circle,inner sep=0mm,minimum size=1.25mm,line width=0mm]

\Crefname{figure}{Fig.}{Figs.}
\crefname{figure}{fig.}{figs.}
\Crefname{tabular}{Tab.}{Tabs.}
\crefname{tabular}{tab.}{tabs.}
\Crefname{section}{Sect.}{Sects.}
\crefname{section}{sect.}{sects.}
\Crefname{equation}{Eq.}{Eqs.}
\crefname{equation}{eq.}{eqs.}
\creflabelformat{equation}{#2#1#3}

\setitemize{noitemsep,topsep=0pt,parsep=0pt,partopsep=0pt,leftmargin=12.0pt}
\setenumerate{noitemsep,topsep=0pt,parsep=0pt,partopsep=0pt,leftmargin=12.5pt}
\setdescription{noitemsep,topsep=0pt,parsep=0pt,partopsep=0pt,leftmargin=12.5pt}

\newcommand{\modest}{\textsc{\mbox{Modest}}\xspace}
\newcommand{\toolset}{\textsc{\mbox{Modest} Toolset}\xspace}
\newcommand{\jani}{\textsc{\mbox{Jani}}\xspace}
\newcommand{\tool}[1]{\textsc{#1}}

\newcommand{\lfp}{\textsf{lfp}\:}
\renewcommand{\max}{\ensuremath{\mathrm{max}}}
\renewcommand{\min}{\ensuremath{\mathrm{min}}}
\newcommand{\opt}{\ensuremath{\mathit{opt}}}
\newcommand{\mcsta}{\textsf{mcsta}\xspace}
\newcommand{\eg}{e.g.\ }
\newcommand{\ie}{i.e.\ }
\newcommand{\etal}{et al.\xspace}
\newcommand{\wrt}{w.r.t.\xspace}
\newcommand{\sunit}[1]{\text{\begin{scriptsize}\,#1\end{scriptsize}}}

\renewcommand{\iff}{\ensuremath{\Leftrightarrow}\xspace}
\newcommand{\set}[1]{\ensuremath{\{\,#1\,\}}}
\newcommand{\tuple}[1]{\ensuremath{\langle #1 \rangle}}
\newcommand{\powerset}[1]{\ensuremath{2^{#1}}\xspace}
\newcommand{\defeq}{\mathrel{\vbox{\offinterlineskip\ialign{\hfil##\hfil\cr{\tiny \rm def}\cr\noalign{\kern0.30ex}$=$\cr}}}}
\newcommand{\Dist}[1]{\ensuremath{\mathit{Dist}({#1})}\xspace}
\newcommand{\support}[1]{\ensuremath{\mathit{spt}({#1})}\xspace}

\newcommand{\RRpluszero}{\ensuremath{\mathbb{R}^+_0}\xspace}

\newcommand{\NN}{\ensuremath{\mathbb{N}}\xspace}

\newcommand{\VV}{\ensuremath{\mathbb{V}}\xspace}
\newcommand{\True}[0]{\ensuremath{\mathit{true}}\xspace}
\newcommand{\False}[0]{\ensuremath{\mathit{false}}\xspace}
\renewcommand{\vec}[1]{\bar{#1}}

\usepackage{listings}
\makeatletter
\lst@AddToHook{OnEmptyLine}{\vspace{-0.75\baselineskip}}
\makeatother

\usepackage[vlined,linesnumbered]{algorithm2e}

\SetAlgoCaptionLayout{raggedright}
\SetAlCapFnt{\small}
\SetAlCapNameFnt{\small}
\SetAlCapSkip{\abovecaptionskip\relax}
\SetEndCharOfAlgoLine{}
\SetArgSty{textrm}
\SetKwInOut{Input}{Input}\SetKwInOut{Output}{Output}
\SetCommentSty{textit}
\SetKw{Break}{break}
\SetKwProg{Type}{type}{}{end}
\SetKwProg{Function}{function}{}{end}
\SetKwFor{ForEach}{foreach}{do}{}
\SetKwFor{WhileTrue}{repeat}{}{end}
\SetKw{Continue}{continue}
\SetKwRepeat{DoWhile}{repeat}{while}
\SetKwProg{Fn}{function}{}{}
\crefname{algocf}{alg.}{algs.}
\Crefname{algocf}{Alg.}{Algs.}

\definecolor{color1}{RGB}{55,126,184}
\definecolor{color2}{RGB}{228,26,28}
\definecolor{color3}{RGB}{77,175,74}
\definecolor{color4}{RGB}{152,78,163}
\definecolor{color5}{RGB}{255,127,0}

\newboolean{hideplots}
\newlength{\scatterplotsize}
\usepackage{pgfplots}
\pgfplotsset{compat=1.14}
\newcommand{\scatterplotiters}[7]{%
\ifthenelse{\boolean{hideplots}}{%
  \begin{minipage}[c][\scatterplotsize][c]{\scatterplotsize}
    \centering
    Scatter plot: #3 vs.\ #5
  \end{minipage}%
}{%
  \begin{tikzpicture}
    \begin{axis}[
      width=#6,
      height=#6,
      axis equal image,
      xmin=20,
      ymin=20,
      xmax=5000,
      ymax=5000,
      xmode=log,
      ymode=log,
      axis x line=bottom,
      axis y line=left,
      xtick={20,200,2000},
      xticklabels={${\leq}\,20$,$200$},
      ytick={20,200,2000},
      yticklabels={${\leq}\,20$,$200$,$2000$},
      xlabel={\strut#3},
      xlabel style={yshift=5pt},
      ylabel={\strut #5},
      ylabel style={yshift=-10pt},
      legend pos=north east,
      legend columns=-1,
      legend style={nodes={scale=0.75, transform shape},inner sep=1.5pt,yshift=16pt,xshift=0.5pt},
      legend cell align={left},
      extra x ticks = 4000,
      extra x tick labels = {${\geq}\,4\,\text{k}$},
      extra x tick style = {grid = major},
      extra y ticks = 4000,
      extra y tick labels = {${\geq}\,4\,\text{k}$},
      extra y tick style = {grid = major},
    ]
    \addplot[
      scatter,
      only marks,
      scatter/classes={
        dtmc={mark=square*,color1,mark size=1.25},
        ctmc={mark=diamond*,color3,mark size=1.75},
        mdp={mark=triangle*,color2,mark size=1.75},
        ma={mark=pentagon*,color4,mark size=1.50},
        pta={mark=*,color5,mark size=1.33},
        dtmc-inc={mark=square,color1,mark size=1.25},
        ctmc-inc={mark=diamond,color3,mark size=1.75},
        mdp-inc={mark=triangle,color2,mark size=1.75},
        ma-inc={mark=pentagon,color4,mark size=1.50},
        pta-inc={color5,mark size=1.33}
      },
      scatter src=explicit symbolic
    ]%
    table [col sep=semicolon,x={#2-iters},y={#4-iters},meta=catcorrect] {#1};
    \ifthenelse{\NOT\equal{#7}{false}}{\legend{DTMC, CTMC, MDP, MA, PTA}}{}
    \addplot[no marks] coordinates {(1,1) (4000,4000)};
    \addplot[no marks, densely dotted] coordinates {(2,1) (4000,2000)};
    \addplot[no marks, densely dotted] coordinates {(1,2) (2000,4000)};
    \end{axis}
  \end{tikzpicture}
}}
\newcommand{\scatterplotpvi}[7]{%
\ifthenelse{\boolean{hideplots}}{%
  \begin{minipage}[c][\scatterplotsize][c]{\scatterplotsize}
    \centering
    Scatter plot: #3 vs.\ #5
  \end{minipage}
}{%
  \begin{tikzpicture}
    \begin{axis}[
      width=#6,
      height=#6,
      axis equal image,
      xmin={1},
      ymin={1},
      xmax=17,
      ymax=17,
      xmode=log,
      ymode=log,
      axis x line=bottom,
      axis y line=left,
      xtick={1,2,4,8},
      xticklabels={${\leq}\,1$,2,4,8},
      ytick={1,2,4,8},
      yticklabels={${\leq}\,1$,2,4,8},
      xlabel={\strut#3},
      xlabel style={yshift=5pt},
      ylabel={\strut #5},
      ylabel style={yshift=-12pt},
      legend pos=north east,
      legend columns=-1,
      legend style={nodes={scale=0.75, transform shape},inner sep=1.5pt,yshift=16pt,xshift=0.5pt},
      legend cell align={left},
      extra x ticks = 15,
      extra x tick labels = {${\geq}\,16$},
      extra x tick style = {grid = major},
      extra y ticks = 15,
      extra y tick labels = {${\geq}\,16$},
      extra y tick style = {grid = major},
    ]
    \addplot[
      scatter,
      only marks,
      scatter/classes={
        dtmc={mark=square*,color1,mark size=1.25},
        ctmc={mark=diamond*,color3,mark size=1.75},
        mdp={mark=triangle*,color2,mark size=1.75},
        ma={mark=pentagon*,color4,mark size=1.50},
        pta={mark=*,color5,mark size=1.33},
        dtmc-inc={mark=square,color1,mark size=1.25},
        ctmc-inc={mark=diamond,color3,mark size=1.75},
        mdp-inc={mark=triangle,color2,mark size=1.75},
        ma-inc={mark=pentagon,color4,mark size=1.50},
        pta-inc={color5,mark size=1.33}
      },
      scatter src=explicit symbolic
    ]%
    table [col sep=semicolon,x={#2-stime},y={#4-stime},meta=catcorrect] {#1};
    \ifthenelse{\NOT\equal{#7}{false}}{\legend{DTMC, CTMC, MDP, MA, PTA}}{}
    \addplot[no marks] coordinates {(1,1) (15,15)};
    \addplot[no marks, densely dotted] coordinates {(2,1) (15,7.5)};
    \addplot[no marks, densely dotted] coordinates {(1,2) (7.5,15)};
    \end{axis}
  \end{tikzpicture}
}}
\newcommand{\scatterplotevi}[7]{%
\ifthenelse{\boolean{hideplots}}{%
  \begin{minipage}[c][\scatterplotsize][c]{\scatterplotsize}
    \centering
    Scatter plot: #3 vs.\ #5
  \end{minipage}
}{%
  \begin{tikzpicture}
    \begin{axis}[
      width=#6,
      height=#6,
      axis equal image,
      xmin={1},
      ymin={1},
      xmax=47,
      ymax=47,
      xmode=log,
      ymode=log,
      axis x line=bottom,
      axis y line=left,
      xtick={1,2,4,8,16,32},
      xticklabels={${\leq}\,1$,$2$,$4$,$8$,$16$},
      ytick={1,2,4,8,16,32},
      yticklabels={${\leq}\,1$,$2$,$4$,$8$,$16$,$32$},
      xlabel={\strut#3},
      xlabel style={yshift=5pt},
      ylabel={\strut #5},
      ylabel style={yshift=-12pt},
      legend pos=north east,
      legend columns=-1,
      legend style={nodes={scale=0.75, transform shape},inner sep=1.5pt,yshift=16pt,xshift=0.5pt},
      legend cell align={left},
      extra x ticks = 40,
      extra x tick labels = {${\geq}\,40$},
      extra x tick style = {grid = major},
      extra y ticks = 40,
      extra y tick labels = {${\geq}\,40$},
      extra y tick style = {grid = major},
    ]
    \addplot[
      scatter,
      only marks,
      scatter/classes={
        dtmc={mark=square*,color1,mark size=1.25},
        ctmc={mark=diamond*,color3,mark size=1.75},
        mdp={mark=triangle*,color2,mark size=1.75},
        ma={mark=pentagon*,color4,mark size=1.50},
        pta={mark=*,color5,mark size=1.33},
        dtmc-inc={mark=square,color1,mark size=1.25},
        ctmc-inc={mark=diamond,color3,mark size=1.75},
        mdp-inc={mark=triangle,color2,mark size=1.75},
        ma-inc={mark=pentagon,color4,mark size=1.50},
        pta-inc={color5,mark size=1.33}
      },
      scatter src=explicit symbolic
    ]%
    table [col sep=semicolon,x={#2-stime},y={#4-stime},meta=catcorrect] {#1};
    \ifthenelse{\NOT\equal{#7}{false}}{\legend{DTMC, CTMC, MDP, MA, PTA}}{}
    \addplot[no marks] coordinates {(1,1) (40,40)};
    \addplot[no marks, densely dotted] coordinates {(2,1) (40,20)};
    \addplot[no marks, densely dotted] coordinates {(1,2) (20,40)};
    \end{axis}
  \end{tikzpicture}
}}
\newcommand{\scatterplotp}[7]{%
\ifthenelse{\boolean{hideplots}}{%
  \begin{minipage}[c][\scatterplotsize][c]{\scatterplotsize}
    \centering
    Scatter plot: #3 vs.\ #5
  \end{minipage}
}{%
  \begin{tikzpicture}
    \begin{axis}[
      width=#6,
      height=#6,
      axis equal image,
      xmin={1},
      ymin={1},
      xmax=17,
      ymax=17,
      xmode=log,
      ymode=log,
      axis x line=bottom,
      axis y line=left,
      xtick={1,2,4,8},
      xticklabels={${\leq}\,1$,$2$,$4$,$8$},
      ytick={1,2,4,8},
      yticklabels={${\leq}\,1$,$2$,$4$,$8$},
      xlabel={\strut#3},
      xlabel style={yshift=5pt},
      ylabel={\strut #5},
      ylabel style={yshift=-12pt},
      legend pos=north east,
      legend columns=-1,
      legend style={nodes={scale=0.75, transform shape},inner sep=1.5pt,yshift=16pt,xshift=0.5pt},
      legend cell align={left},
      extra x ticks = 16,
      extra x tick labels = {${\geq}\,16$},
      extra x tick style = {grid = major},
      extra y ticks = 16,
      extra y tick labels = {${\geq}\,16$},
      extra y tick style = {grid = major},
    ]
    \addplot[
      scatter,
      only marks,
      scatter/classes={
        dtmc={mark=square*,color1,mark size=1.25},
        ctmc={mark=diamond*,color3,mark size=1.75},
        mdp={mark=triangle*,color2,mark size=1.75},
        ma={mark=pentagon*,color4,mark size=1.50},
        pta={mark=*,color5,mark size=1.33}
      },
      scatter src=explicit symbolic
    ]%
    table [col sep=semicolon,x={#2-stime},y={#4-stime},meta=category] {#1};
    \ifthenelse{\NOT\equal{#7}{false}}{\legend{DTMC, CTMC, MDP, MA, PTA}}{}
    \addplot[no marks] coordinates {(1,1) (16,16)};
    \addplot[no marks, densely dotted] coordinates {(2,1) (16,8)};
    \addplot[no marks, densely dotted] coordinates {(1,2) (8,16)};
    \end{axis}
  \end{tikzpicture}
}}
\newcommand{\scatterplote}[7]{%
\ifthenelse{\boolean{hideplots}}{%
  \begin{minipage}[c][\scatterplotsize][c]{\scatterplotsize}
    \centering
    Scatter plot: #3 vs.\ #5
  \end{minipage}
}{%
  \begin{tikzpicture}
    \begin{axis}[
      width=#6,
      height=#6,
      axis equal image,
      xmin={1},
      ymin={1},
      xmax=70,
      ymax=70,
      xmode=log,
      ymode=log,
      axis x line=bottom,
      axis y line=left,
      xtick={1,2,4,8,16,32},
      xticklabels={${\leq}\,1$,$2$,$4$,$8$,$16$,$32$},
      ytick={1,2,4,8,16,32},
      yticklabels={${\leq}\,1$,$2$,$4$,$8$,$16$,$32$},
      xlabel={\strut#3},
      xlabel style={yshift=5pt},
      ylabel={\strut #5},
      ylabel style={yshift=-12pt},
      legend pos=north east,
      legend columns=-1,
      legend style={nodes={scale=0.75, transform shape},inner sep=1.5pt,yshift=16pt,xshift=0.5pt},
      legend cell align={left},
      extra x ticks = 60,
      extra x tick labels = {${\geq}\,60$},
      extra x tick style = {grid = major},
      extra y ticks = 60,
      extra y tick labels = {${\geq}\,60$},
      extra y tick style = {grid = major},
    ]
    \addplot[
      scatter,
      only marks,
      scatter/classes={
        dtmc={mark=square*,color1,mark size=1.25},
        ctmc={mark=diamond*,color3,mark size=1.75},
        mdp={mark=triangle*,color2,mark size=1.75},
        ma={mark=pentagon*,color4,mark size=1.50},
        pta={mark=*,color5,mark size=1.33}
      },
      scatter src=explicit symbolic
    ]%
    table [col sep=semicolon,x={#2-stime},y={#4-stime},meta=category] {#1};
    \ifthenelse{\NOT\equal{#7}{false}}{\legend{DTMC, CTMC, MDP, MA, PTA}}{}
    \addplot[no marks] coordinates {(1,1) (60,60)};
    \addplot[no marks, densely dotted] coordinates {(2,1) (60,30)};
    \addplot[no marks, densely dotted] coordinates {(1,2) (30,60)};
    \end{axis}
  \end{tikzpicture}
}}

\hypersetup{
  pdftitle = {Optimistic Value Iteration}
}
\begin{document}

\title{%
Optimistic Value Iteration%
\thanks{%
The authors are listed alphabetically.
This work was supported by
ERC Advanced Grant 787914 ({\scriptsize FRAPPANT}), DFG Research Training Group 2236 ({\scriptsize UnRAVeL}), and NWO VENI grant no.\ 639.021.754.}%
}
\author{
Arnd Hartmanns\inst{1}
\and Benjamin Lucien Kaminski\inst{2}
}

\institute{
University of Twente, Enschede, The Netherlands
\and RWTH Aachen University, Aachen, Germany}
\date{\today}
\maketitle

\begin{abstract}
Markov decision processes are widely used for planning and verification in settings that combine controllable or adversarial choices with probabilistic behaviour.
The standard analysis algorithm, value iteration, only provides lower bounds on infinite-horizon probabilities and rewards.
Two ``sound'' variations, which also deliver an upper bound, have recently appeared.
In this paper, we present a new sound approach that leverages value iteration's ability to \emph{usually} deliver \emph{tight} lower bounds:
we obtain a lower bound via standard value iteration, use the result to ``guess'' an upper bound, and prove the latter's correctness.
The approach is \emph{easy} to implement, does not require extra precomputations or a priori state space transformations, and works for computing reachability probabilities as well as expected rewards.
It is also \emph{fast}, as we show via an extensive experimental evaluation using our publicly available implementation within the \mcsta model checker of the \toolset.
\end{abstract}

\section{Introduction}
\label{sec:Introduction}

Markov decision processes (MDP,~\cite{Put94}) are a widely-used formalism to represent discrete-state and -time systems in which \emph{probabilistic} effects meet controllable \emph{nondeterministic} decisions.
The former may arise from an environment or agent whose behaviour is only known statistically (\eg message loss in wireless communication or statistical user profiles), or it may be intentional as part of a randomised algorithm (such as exponential backoff in Ethernet).
The latter may be under the control of the system---then we are in a planning setting and typically look for a \emph{scheduler} (or strategy, policy) that minimises the probability of unsafe behaviour or maximises a reward---or it may be considered adversarial, which is the standard assumption in verification: we want to establish that the maximum probability of unsafe behaviour is below, or that the minimum reward is above, a specified threshold.
Extensions of MDP cover continuous time~\cite{EHZ10,KNSS02}, and the analysis of complex formalisms such as stochastic hybrid automata~\cite{FHHWZ11} can be reduced to the analysis of MDP abstractions.

The standard algorithm to compute optimal (maximum or minimum) probabilities or reward values on MDP is \emph{value iteration} (VI).
It implicitly computes the corresponding optimal scheduler, too.
It keeps track of a value for every state of the MDP, locally improves the values iteratively until a ``convergence'' criterion is met, and then reports the final value for the initial state as the overall result.
The initial values are chosen to be an underapproximation of the true values (\eg 0 for all states in case of probabilities or non-negative rewards).
The final values are then an improved underapproximation of the true values.
For unbounded (infinite-horizon) properties, there is unfortunately no (known) convergence criterion that could guarantee a predefined error on the final result.
Still, probabilistic model checkers such as \tool{Prism}~\cite{KNP11} report the final result obtained via simple relative or absolute global error criteria as the definitive probability.
This is because, on \emph{most} case studies considered so far, value iteration in fact converges fast enough that the (relative or absolute) difference between the reported and the true value meets the error $\epsilon$ specified for the convergence criterion.
Only relatively recently has this problem of soundness come to the attention of the probabilistic verification and planning communities~\cite{BCCFKKPU14,HM14,MLG05}.
First highlighted on hand-crafted counterexamples, it has by now been found to affect benchmarks and real-life case studies, too~\cite{BKLPW17}.

The first proposal to compute sound reachability probabilities was to use \emph{interval iteration} (II)~\cite{HM18}.
The idea is to perform two value iterations concurrently, one starting from 0 as before, and one starting from 1 for all relevant states.
The latter iterations improve an overapproximation of the true values, and the entire process can be stopped once the (relative or absolute) difference between the two values for the initial state is below the specified $\epsilon$.
Interval iteration, however, requires the MDP to be in a form where value iteration has a single fixed point.
For minimum probabilities, this is achieved via graph-based (\ie not numerical) precomputations~\cite[algs.\ 1-4]{FKNP11}.
For maximum probabilities, however, additionally end components need to be eliminated, requiring a state space transformation whose extra memory usage exacerbates the state space explosion problem.
Baier \etal extended interval iteration to expected accumulated reward values~\cite{BKLPW17}; here, the complication is to find initial values that are guaranteed to be an overapproximation.
The proposed graph-based algorithm in practice computes very conservative initial values, from which many iterations are needed until convergence.
More recently, \emph{sound value iteration} (SVI)~\cite{QK18} improved upon interval iteration by computing upper bounds on-the-fly and performing larger value improvements per iteration, for both probabilities and expected rewards.
It still requires the same precomputations and end component reduction as interval iteration, though; it only does not need a priori upper bounds for expected rewards (although they may improve performance if provided).

\paragraph{Our contribution.}
We present (in \Cref{sec:OVI}) a new approach to computing sound reachability probabilities and expected rewards that is both simple and practically efficient.
We first (1)~perform standard value iteration until ``convergence'', resulting in a lower bound on the value for every state.
To this we (2)~apply heuristics to ``guess'', for every state, a candidate upper bound value.
Further iterations (3)~then confirm (if all values decrease) or disprove (if all values increase, or lower and upper bounds cross) the soundness of the upper bounds.
In the latter case, we perform more lower bound iterations with reduced $\epsilon$ before retrying from step~2.
In problematic cases, many retries may be needed, and performance may be worse than interval or sound value iteration.
However, on the vast majority of existing case studies, value iteration already worked well, and our approach attaches a soundness proof to its result with moderate overhead.
We thus refer to it as \emph{optimistic value iteration} (OVI).
It does not require any of the precomputations, end component reductions, or a priori bound computations that are needed for II and SVI, further simplifying implementations and improving scalability.
Our experimental evaluation in \Cref{sec:Experiments} uses all applicable models from the Quantitative Verification Benchmark Set~\cite{HKPQR19} to confirm that OVI indeed performs as expected. 
It uses our publicly available implementations of II, SVI, and now OVI in the \mcsta model checker of the \toolset~\cite{HH14}.

\paragraph{Related work.}
As an alternative to the iterative numeric road that we take in this paper, guaranteed correct results (modulo implementation errors) can also be obtained by using precise rational arithmetic.
It does not combine too well with sound iterative methods like II or SVI due to the increasingly small differences between the values and the actual solution.
The probabilistic model checker \tool{Storm}~\cite{DJK017} thus combines topological decomposition, policy iteration, and exact solvers for linear equation systems based on Gaussian elimination when asked to use rational arithmetic~\cite[Section 7.4.8]{Hen18}.
Alternatively, rational search~\cite{BMCSV17} can be used, which proceeds in a very similar way to OVI:
starting from the (floating-point) result of value iteration, it finds the next rational number, checks whether it is the correct result, and if not, continues the floating-point iterations.
The disadvantage of both of these methods is the significant runtime cost for performing the unlimited-precision calculations, limiting them to relatively smaller~MDP.

\section{Preliminaries}
\label{sec:Preliminaries}

$\RRpluszero$ is the set of all non-negative real numbers.
We write $\set{ x_1 \mapsto y_1, \dots }$ to denote the function that maps all $x_i$ to $y_i$, and if necessary in the respective context, implicitly maps to~$0$ all $x$ for which no explicit mapping is specified.
Given a set $S$, its powerset is $\powerset{S}$.
A (discrete) \emph{probability distribution} over $S$ is a function $\mu \in S \to [0, 1]$ with countable \emph{support} $\support{\mu} \defeq \set{ s \in S \mid \mu(s) > 0 }$ and $\sum_{s \in \support{\mu}} \mu(s) = 1$.
$\Dist{S}$ is the set of all probability distributions over $S$. 

\paragraph{Markov decision processes}
(MDP) combine nondeterministic choices as in labelled transition systems with discrete probabilistic decisions as in discrete-time Markov chains (DTMC).
We define them formally and describe their semantics.

\begin{definition}
\label{def:MDP}
A \emph{Markov decision process} (MDP) is a triple\\[5pt]
\centerline{
$M =\tuple{S, s_I, T}$
}\\[-5pt]
where
\begin{itemize}
\item
$S$ is a finite set of \emph{states} with \emph{initial state} $s_I \in S$ and
\item
$T \colon \mathit{S} \to \powerset{\Dist{\RRpluszero \times S}}$ is the \emph{transition function}.
\end{itemize}
$T(s)$ must be finite and non-empty for all $s \in S$.
\end{definition}
For $s \in S$, an element of $T(s)$ is a \emph{transition}, and a pair $\tuple{r, s'} \in \support{T(s)}$ is a \emph{branch} to successor state $s'$ with \emph{reward} $r$ and probability $T(s)(\tuple{r, s'})$.
Let $M^{(s_I')}$ be $M$ but with initial state $s_I'$, and $M^0$ be $M$ with all rewards set to zero.

\begin{example}
\label{ex:MDP}
\begin{figure}[t]
\begin{floatrow}
\ffigbox[0.26\textwidth]{
\begin{tikzpicture}[on grid,auto]
  \node[state] (s0) {$\mathstrut s_0$};
  \coordinate[above=0.3 of s0.north] (start);
  \node[] (me) [above left=0.5 and 0.8 of s0] {\small$M_e$:};
  \node[dot] (n0) [below right=0.5 and 0.5 of s0] {};
  \node[state] (sp) [below=1.25 of s0] {$\mathstrut s_+$};
  \node[state] (sm) [below right=1.25 and 1.0 of s0] {$\mathstrut s_-$};
  \node[state] (s1) [below=1.5 of sp] {$\mathstrut s_1$};
  \node[state] (s2) [below=1.5 of sm] {$\mathstrut s_2$};
  \node[dot] (n1) [above left=0.75 and 0.5 of s2] {};
  ;
  \path[-]
    (s0) edge[bend left] node[inner sep=1pt] {\texttt{a}} (n0)
    (s2) edge[bend right=20] node[inner sep=1.5pt,pos=0.1,swap] {\texttt{c}} (n1)
  ;
  \path[->]
    (start) edge node {} (s0)
    (s0) edge [bend right=60] node[swap,pos=0.06,inner sep=1pt] {\texttt{b}} (s1)
    (s1) edge [bend left] node {} (s2)
    (s2) edge [bend left] node {} (s1)
    (n0) edge[bend left] node[align=left,inner sep=-2pt,pos=0.8] {$0.1$\\[-1pt]$~~{+1}$} (sm)
    (n0) edge[bend right] node[align=right,swap,inner sep=-2pt,pos=0.8] {$0.1$\\[-1pt]\phantom{${+1}~~$}} (sp)
    (n0) edge[out=20,in=30,looseness=3] node[align=left,right,inner sep=3pt,pos=0.45] {$0.8$\\[-1pt]${+1}$} (s0)
    (n1) edge[bend right] node[align=left,swap,inner sep=-2pt,pos=0.7] {$~~0.6$\\[-1pt]${+1}$} (sm)
    (n1) edge[bend left] node[align=right,inner sep=-2pt,pos=0.7] {$0.4~~$\\[-1pt]\phantom{${+1}$}} (sp)
    (sm) edge [loop,out=30,in=-30,looseness=5] node {} (sm)
    (sp) edge [loop,out=30,in=-30,looseness=5] node {} (sp)
  ;
\end{tikzpicture}}{
\caption{Example MDP}
\label{fig:ExampleMDP}}%
\capbtabbox[0.71\textwidth]{%
\renewcommand{\arraystretch}{1.05}
\setlength{\tabcolsep}{2.0pt}
\scriptsize
\begin{tabular}{@{}cllllllll@{}}
\toprule
$i$ & $v(s_0)$ & $u(s_0)$ & $v(s_1)$ & $u(s_1)$ & $v(s_2)$ & $u(s_2)$ & $\mathit{error}$ & $\epsilon_\mathit{VI}$ \\
\midrule
$0$ & $0$ & & $0$ & & $0$ & & & $0.05$ \\
$1$ & $0.1$ & & $0$ & & $0.4$ & & $0.4$ & $0.05$ \\
$2$ & $0.18$ & & $0.4$ & & $0.4$ & & $0.4$ & $0.05$ \\
$3$ & $0.4$ & & $0.4$ & & $0.4$ & & $0.22$ & $0.05$ \\
$4$ & $0.42$ & $\textsl{0.47}$ & $0.4$ & $\textsl{0.45}$ & $0.4$ & $\textsl{0.45}$ & $0.02$ & $0.05$ \\
$5$ & $0.436$ & $0.476$ & $0.4$ & $0.45$ & $0.4$ & $0.45$ & $0.016$ & \\
$6$ & $0.4488$ & & $0.4$ & & $0.4$ & & $0.0128$ & $0.008$ \\
$7$ & $0.45904$ & & $0.4$ & & $0.4$ & & $0.01024$ & $0.008$ \\
$8$ & $0.467232$ & & $0.4$ & & $0.4$ & & $0.008192$ & $0.008$ \\
$9$ & $0.4737856$ & $\textsl{0.5237856}$ & $0.4$ & $\textsl{0.45}$ & $0.4$ & $\textsl{0.45}$ & $0.0065536$ & $0.008$ \\
$10$ & $0.47902848$ & $0.51902848$ & $0.4$ & $0.45$ & $0.4$ & $0.45$ & $0.00524288$ & \\
\bottomrule
\end{tabular}}{
\caption{VI and OVI example on $M_e$}
\label{tab:ExampleOVI}
}
\end{floatrow}
\end{figure}
\Cref{fig:ExampleMDP} shows our example MDP $M_e$.
We draw transitions as lines to an intermediate node from which branches labelled with probability and reward (if not zero) lead to successor states.
We omit the intermediate node and probability~$1$ for transitions with a single branch, and label some transitions to refer to them in the text.
$M^e$ has 5~states, 5~transitions, and 8~branches.
\end{example}
In practice, higher-level modelling languages like \modest~\cite{HHHK13} are used to specify MDP.
The semantics of an MDP is captured by its \emph{paths}.
A path represents a concrete resolution of all nondeterministic and probabilistic choices.
Formally:

\begin{definition}
A \emph{finite path} is a sequence\\[5pt]
\centerline{%
$\pi_\mathrm{fin} = s_0\, \mu_0\, r_0\, s_1\, \mu_1\, r_1\, s_2 \dots  \mu_{n-1} r_{n-1} s_n$}\\[5pt]
where $s_i \in S$ for all $i \in \set{ 0, \dots, n }$ and $\exists\, \mu_i \in T(s_i) \colon \tuple{r_i, s_{i+1}} \in \support{\mu_i}$ for all $i \in \set{ 0, \dots, n - 1 }$.
Let $|\pi_\mathrm{fin}| \defeq n$, $\mathrm{last}({\pi_\mathrm{fin}}) \defeq s_n$, and $\mathrm{rew}({\pi_\mathrm{fin}}) = \sum_{i=0}^{n-1} r_i$.
$\Pi_\mathit{fin}$ is the set of all finite paths starting in $s_I$.
A \emph{path} is an analogous infinite sequence $\pi$, and $\Pi$ are all paths starting in $s_I$.
We define $s \in \pi \iff \exists\, i \colon s = s_i$, and $\pi_{\to G}$ as the shortest prefix of $\pi$ that contains a state in $G \subseteq S$, or $\bot$ if $\pi$ contains no such state.
Let $\mathrm{rew}(\bot) \defeq \infty$.%
\end{definition}
A scheduler (or \emph{adversary}, \emph{policy} or \emph{strategy}) only resolves the nondeterministic choices of~$M$.
For this paper, memoryless deterministic schedulers suffice.
\begin{definition}
\label{def:MDPReductionFunction}
A function $\mathfrak{s} \colon S \to \Dist{\RRpluszero \times S}$ is a \emph{scheduler} if, for all $s \in S$, we have $\mathfrak{s}(s) \in T(s)$.
The set of all schedulers of~$M$ is $\mathfrak{S}(M)$.
\end{definition}
Given an MDP $M$ as above, let $M|_\mathfrak{s} = \tuple{S, s_I, T|_\mathfrak{s}}$ with $T|_\mathfrak{s}(s) = \set{\mathfrak{s}(s)}$ be the DTMC induced by $\mathfrak{s}$.
Via the standard cylinder set construction~\cite[Sect.\ 2.2]{FKNP11} on $M|_\mathfrak{s}$, a scheduler induces a probability measure $\mathbb{P}_\mathfrak{s}^M$ on measurable sets of paths starting in $s_I$.
For goal state $g \in S$, the maximum and minimum probabilities of reaching $g$ are defined as $\mathrm{P}_{\!\max}^M(\diamond\: g) = \sup_{\mathfrak{s} \in \mathfrak{S}} \mathbb{P}_\mathfrak{s}^M(\set{ \pi \in \Pi \mid g \in \pi })$ and $\mathrm{P}_{\!\min}^M(\diamond\: g) = \inf_{\mathfrak{s} \in \mathfrak{S}} \mathbb{P}_\mathfrak{s}^M(\set{ \pi \in \Pi \mid g \in \pi })$, respectively.
The definition extends to sets $G$ of goal states.
Let $R_G^M \colon \Pi \to \RRpluszero$ be the random variable defined by $R_G^M(\pi) = \mathrm{rew}(\pi_{\to G})$ and let $\mathbb{E}_\mathfrak{s}^M(G)$ be the expected value of $R_G^M$ under $\mathbb{P}_\mathfrak{s}^M$.
Then the maximum and minimum expected reward to reach $G$ is defined as $\mathrm{E}_\max^M(G) = \sup_{\mathfrak{s}}\mathbb{E}_\mathfrak{s}^M(G)$ and $\mathrm{E}_\min^M(G) = \inf_{\mathfrak{s}}\mathbb{E}_\mathfrak{s}^M(G)$, respectively.
We omit the superscripts for $M$ when they are clear from the context.
From now on, whenever we have an MDP with a set of goal states $G$, we assume that they have been made absorbing, \ie for all $g \in G$ we only have a self-loop: $T(g) = \set{ \set{ \tuple{0, g} \mapsto 1 } }$.

\begin{definition}
An \emph{end component} of $M$ as above is a (sub-)MDP $\tuple{S', T', s_I'}$ where $S' \subseteq S$, $T'(s) \subseteq T(s)$ for all $s \in S'$, if $\mu \in T'(s)$ for some $s \in S'$ and $\tuple{r, s'} \in \support{\mu}$ then $r = 0$, and the directed graph with vertex set $S'$ and edge set $\set{ \tuple{s, s'} \mid \exists\,\mu \in T'(s) \colon \tuple{0, s'} \in \support{\mu} }$ is strongly connected.
\end{definition}

\section{Value Iteration}

The standard algorithm to compute reachability probabilities and expected rewards is \emph{value iteration} (VI)~\cite{Put94}.
In this section, we recall its theoretical foundations as well as its limitations regarding convergence.

Let $\VV = \set{v ~|~ v \colon S \to \RRpluszero \cup \{\infty\}}$ be a space of vectors of values.
It can easily be shown that $\tuple{\VV,\, {\preceq}}$, with\\[5pt]
\centerline{
$v \preceq w \qquad\text{if and only if}\qquad \forall\, s \in S\colon v(s) \leq w(s)$,
}\\[5pt]
forms a complete lattice, i.e.\ every subset $V \subseteq \VV$ has a supremum (and an infimum) in $\VV$ with respect to $\preceq$.

Minimum and maximum reachability probabilities and expected rewards can be expressed as the \emph{least fixed point} of the \emph{Bellman operator}~\mbox{$\Phi\colon \VV \to \VV$} given by%
\vspace{-6pt}%
\begin{align*}
\Phi(v) \defeq \lambda\: s.~ \begin{cases}
        \opt_{\mu \in T(s)} ~ \sum_{\tuple{r, s'} \in \support{\mu}} ~ \mu(s') \cdot (r + v(s')) & \text{ if } s \in S_?\\
        d  &\text{ if } s \not\in S_?~,
\end{cases} 
\end{align*}\\[-11pt]%
where $\opt \in \set{ \max, \min }$ and the choice of both $S_? \subseteq S$ and $d$ depends on whether we wish to compute reachability probabilities or expected rewards.
In any case, the Bellman operator $\Phi$ can be shown to be Scott-continuous \cite{abramskyjung94}, \ie in our case: for any subset $V \subseteq \VV$, we have $\Phi( \sup V) = \sup \Phi(V)$.

The Kleene fixed point theorem for Scott-continuous self-maps on complete lattices \cite{abramskyjung94,DBLP:journals/ipl/LassezNS82} guarantees that the least fixed point of $\Phi$, denoted by $\lfp \Phi$, indeed exists.
Note that $\Phi$ can have more than one fixed point---only the least fixed point is guaranteed to exists (and is necessarily unique).
In addition to mere existence of $\lfp \Phi$, the Kleene fixed point theorem states that $\lfp \Phi$ can \mbox{be expressed by}%
\begin{align}
        \lfp \Phi = \lim_{n \to \infty} \Phi^n(\vec{0}) \label{eq:limit-lfp}
\end{align}%
where $\vec{0} \in \VV$ is the zero vector and $\Phi^n(v)$ denotes $n$-fold application of $\Phi$ to $v$.
\Cref{eq:limit-lfp} forms the theoretical basis for VI:
the algorithm iteratively constructs a sequence of vectors with\\[5pt]
\centerline{
$v_0 = \vec{0} \qquad\text{and}\qquad v_{i+1} = \Phi(v_i)$,
}\\[5pt]
which converges to the sought-after least fixed point.
This convergence is \emph{monotonic}:
for every $n \in \NN$, we have $\Phi^n(\vec{0}) \preceq \Phi^{n+1}(\vec{0})$ and hence $\Phi^n(\vec{0}) \preceq \lfp \Phi$. 
In particular, $\Phi^n(\vec{0})(s_I)$ is an \emph{under}approximation of the sought-after quantity for every $n$.
Note that iterating $\Phi$ on \emph{any} underapproximation $v \preceq \lfp \Phi$ (instead of~$\vec{0}$) will still converge to $\lfp \Phi$ and $\Phi^n(v) \preceq \lfp \Phi$ will hold for any $n$.

For determining concrete reachability probabilities, we operate on $M^0$ and choose $S_? = S \setminus G$ and $d = 1$.
Then the least fixed point of the corresponding Bellman operator satisfies\\[-0pt]
\centerline{
$(\lfp \Phi)(s) = \mathrm{P}_\mathit{\!\!opt}^{M^{(s)}}(\diamond\: G)$,
}\\[5pt]
and VI will iteratively approximate this quantity \emph{from below}.

For determining the expected reward $\mathrm{E}_\opt^{M^{(s)}}(G)$, we operate on $M$ and first have to determine the set $S_\infty$ of states from which the minimum (if $\opt = \max$) or maximum (if $\opt = \min$) probability to reach $G$ is less than $1$.%
\footnote{This can be done via Algs.\ 2 and~4 of \cite{FKNP11}, respectively.
These algorithms do not consider the actual probabilities, but only whether there is a transition and branch (with positive probability) from one state to another or not.
We thus call them \emph{graph-based} (as opposed to \emph{numeric}) algorithms.}
If $s_I \in S_\infty$, then the result is $\infty$ due to the definition of $\mathrm{rew}(\bot)$.
Otherwise, we choose $S_? = S \setminus S_\infty$ and $d = \infty$.
Then, for $\opt = \max$, the least fixed point of the corresponding Bellman operator satisfies\\[5pt]
\centerline{
$(\lfp \Phi)(s) = \mathrm{E}_\opt^{M^{(s)}}(G)$.
}\\[5pt]
Again, VI underapproximates this quantity.
The same holds for $\opt = \min$ if $M$ does not have end components containing states other than those in $G$ and $S_\infty$.

\paragraph{Gauss-Seidel value iteration.}
\begin{algorithm}[t]
\Function{\texttt{GSVI}$(M = \tuple{S, s_I, T}$, $S_?$, $\opt \in \set{ \max, \min }$, $v$, $\epsilon_\mathit{VI})$}{
  \Repeat{$\mathit{error} < \epsilon_\mathit{VI}$\label{alg:VI:Until}}{
    $\mathit{error} := 0$\;
    \ForEach{$s \in S_?$\label{alg:VI:Foreach}}{
      $v_\mathit{new} := \opt_{\mu \in T(s)} \sum_{\tuple{r, s'} \in \support{\mu}}{\mu(s') \cdot (r + v(s'))}$\label{alg:VI:Update}\;
      \lIf{$v_\mathit{new} > 0$}{%
        $\mathit{error} := \max\,\set{\mathit{error}, (v_\mathit{new} - v(s)) / v_\mathit{new} }$\label{alg:VI:Error}
      }
      $v(s) := v_\mathit{new}$
    }
  }
}
\caption{Gauss-Seidel value iteration with relative-error convergence}
\label{alg:VI}
\end{algorithm}%
\Cref{alg:VI} shows the pseudocode of a VI implementation that uses the so-called \emph{Gauss-Seidel optimisation}:
Whereas standard VI needs to store two vectors $v_i$ and $v_{i+1}$, Gauss-Seidel VI stores only a single vector~$v$ and performs updates in place.
This does not affect the correctness of VI, but may speed up convergence depending on the order in which the loop in line~\ref{alg:VI:Foreach} considers the states in $S_?$.
To move towards the least fixed point, we call $\texttt{GSVI}$ with a trivial underapproximation:
$v = \set{ s \mapsto 0 \mid s \in S \setminus G } \cup \set{ s \mapsto 1 \mid s \in G }$ for $\mathrm{P}_\mathit{\!\!opt}(\diamond\: G)$ (and we operate on $M^0$ instead of $M$), and $v = \set{ s \mapsto 0 \mid s \in S \setminus S_\infty } \cup \set{ s \mapsto \infty \mid s \in S_\infty }$ for $\mathrm{E}_\opt(G)$.

\paragraph{Convergence.}
\texttt{GSVI} will not, in general, reach a fixed point (and neither will classical VI); we thus use the standard \emph{relative error} convergence criterion to decide when to stop iterations (lines \ref{alg:VI:Error} and~\ref{alg:VI:Until}).
To use the absolute error, replace line~\ref{alg:VI:Error} by $\mathit{error} := \max\,\set{\mathit{error}, v_\mathit{new} - v(s) }$.
Upon termination of \texttt{GSVI}, $v$ is closer to the least fixed point, but remains an underapproximation.
In particular, the parameter $\epsilon_\mathit{VI}$ (which is $10^{-6}$ by default in most probabilistic model checkers) has, in general, no formal relation whatsoever to the final difference between $v(s_I)$ and $\mathrm{P}_\mathit{\!\!opt}(\diamond\: G)$ or $\mathrm{E}_\opt(G)$, respectively.

\begin{example}
Consider MDP $M_e$ of \Cref{fig:ExampleMDP} again.
The first four rows in the body of \Cref{tab:ExampleOVI} show the values for $v$ after the $i$-th iteration of the outer loop of a call to $\texttt{GSVI}(M_e^0, \set{ s_0, s_1, s_2 }, \max, \set{ s_+ \mapsto 1 } \cup \set{ s \mapsto 0 \mid s \neq s_+ }, 0.05 )$ using absolute-error convergence.
After the fourth iteration, \texttt{GSVI} terminates since the error is less than~$\epsilon_\mathit{VI} = 0.05$; at this point, we have $\mathrm{P}_{\!\max}(\diamond\: s_+) - v(s_0) = 0.08 > \epsilon_\mathit{VI}$.

In $M_e$, states $s_1$ and $s_2$ and the two transitions in between form an end component.
$v = \set{ s \mapsto 1 }$ is another fixed point for the corresponding Bellman operator here; in fact, with appropriate values for $s_1$ and $s_2$, we can obtain fixed points with any $v(s_0) > 0.5$ of our choice.
Similarly, we have $\mathrm{E}_\min^M(\set{s_+, s_-}) = 0.6$ (by scheduling \texttt{b} in $s_0$), but due to the end component (which has only zero-reward transitions by definition), the fixed point is such that $v(s_0) = 0$.
\end{example}
Value iteration thus comes with the two problems of convergence and uniqueness of fixed points.
In practice, the latter is not critical for $\mathrm{P}_{\!\min}$, $\mathrm{P}_{\!\max}$, and $\mathrm{E}_\max$:
we simply call \texttt{GSVI} with a (trivial) underapproximation.
For $\mathrm{E}_\min$, (zero-reward) end components rarely occur in case studies since they indicate Zeno behaviour \wrt to the reward.
As rewards are often associated to time progress, such behaviour would be unrealistic.
To make the fixed points unique, for $\mathrm{E}_\max$ we fix the value of all goal states to~$0$.
For $\mathrm{P}_{\!\min}$, we precompute the set of states that reach the goal with probability~$0$ using algs.\ 1 and~3 of \cite{FKNP11}, then fix their values to~$0$.
For $\mathrm{P}_{\!\max}$ and $\mathrm{E}_\min$, we additionally need to \emph{eliminate end components}:
we determine the maximal end components using algorithms similar to \cite[Alg.\ 1]{HM18}, then replace each of them by a single state, keeping all transitions leading out of the end component.
In contrast to the precomputations, end component elimination changes the structure of the MDP and is thus more memory-intensive, yet a sound probabilistic model checker cannot avoid it for $\mathrm{E}_\min$ properties.

The convergence problem is more severe. 
Current solutions consist of computing an upper bound in addition to the lower bound provided by VI.
Interval iteration (II)~\cite{HM14,BKLPW17} does so by essentially performing, in parallel, a second value iteration on a second vector $u$ that starts from an overapproximation of the values.
For probabilities, the one vector $\vec{1} = \set{s \mapsto 1}$ is a trivial overapproximation; for rewards, more involved graph-based algorithms as presented in~\cite{BKLPW17} need to be used to precompute (a very conservative) one.
Interval iteration terminates as soon as $u(s_I) - v(s_I) \leq 2\epsilon \cdot v(s_I)$ (assuming $\epsilon$ specifies a relative-width requirement; if it is an absolute width, we compare with just $2\epsilon$) and returns $v_{\mathit{II}} = \frac{1}{2}(u(s_I) + v(s_I))$.
With $v_\True = \mathrm{P}_\mathit{\!\!opt}(\diamond\: G)$, it thus guarantees that $v_{\mathit{II}} \in [v_\True - \epsilon \cdot v_\True, v_\True + \epsilon \cdot v_\True]$ and analogously for expected rewards.
To ensure termination, II requires a unique fixed point: the precomputations, and in particular end component elimination for $\mathrm{P}_{\!\max}$, are thus no longer optional.
Sound value iteration (SVI)~\cite{QK18} uses a different approach to deriving upper bounds that makes it perform better overall, and that eliminates the need to precompute an initial overapproximation for expected rewards.
It still requires unique fixed points and hence precomputations.

\section{Optimistic Value Iteration}
\label{sec:OVI}

We now describe a new, practical approach to solving the convergence problem for unbounded reachability and expected rewards.
It exploits the observation that VI does deliver results that are in fact $\epsilon$-close to the true value on \emph{most} case studies to which probabilistic model checking has been applied so far---it only lacks the ability to prove it.
Our approach, called \emph{optimistic value iteration} (OVI), extends standard value iteration with the ability to deliver such a proof.

The key idea is to exploit a property of the Bellman operator $\Phi$ as well as of Gauss-Seidel value iteration as in \Cref{alg:VI} to determine whether a candidate vector is a lower bound, an upper bound, or neither.
The foundation of our approach is basic domain theory:
By Scott-continuity of $\Phi$ it follows easily that $\Phi$ is monotonic, meaning $v \preceq w$ implies $\Phi(v) \preceq \Phi(w)$.
A principle called \emph{Park induction} \cite{park1969fixpoint} for monotonic self-maps on complete lattices yields the following induction rule:
For any $u \in \VV$,%
\vspace{-6pt}\begin{align}
        \Phi(u) \preceq u \qquad\text{implies}\qquad \lfp \Phi \preceq u \label{eq:park}
\end{align}\\[-19pt]
Thus, if we can construct a candidate vector $u$ that satisfies $\Phi(u) \preceq u$, then $u$ is in fact an upper bound on the sought-after least fixed point.
We call such a $u$ an \emph{inductive upper bound}.
Optimistic value iteration uses this insight and can---in a nutshell---be summarised as follows:\\[-5pt]

\fbox{\parbox{0.9\textwidth}{%
\begin{enumerate}
        \item Perform Gauss-Seidel value iteration until the current underapproximation $v$ satisfies the VI convergence criterion.
        \item Heuristically determine a candidate $u$ and compute $\Phi(u)$.
        \item If $\Phi(u) \preceq u$, then $v \preceq \lfp \Phi \preceq u$.
        \begin{itemize}
                \item If $u(s_I) - v(s_I) < 2\epsilon$, \textbf{terminate and return $\boldsymbol{\tfrac{1}{2}\bigl(u(s_I) + v(s_I)\bigr)}$}.
        \end{itemize}
        \item Tweak parameters pertaining to convergence of VI and goto step 1.
\end{enumerate}}}\\[1pt]
\begin{algorithm}[t]
\Function{\texttt{OVI}$(M = \tuple{S, s_I, T}$, $S_?$, $\opt \in \set{ \max, \min }$, $v$, $\epsilon)$}{
  $\mathit{error} := \epsilon$\;
  \WhileTrue{}{
    $\texttt{GSVI}(M, S_?, \opt, v, \mathit{error})$\label{alg:OVI:CallVI}\tcp*{perform standard value iteration}
    $u := \set{ s \mapsto v(s) \cdot (1 + \epsilon) \mid s \in S_? }$\label{alg:OVI:Guess}\tcp*{guess candidate upper bound}
    \WhileTrue{\label{alg:OVI:Verif}\tcp*[f]{start the verification phase}}{
      $\mathit{error} := 0$, $\mathit{up} := \True$, $\mathit{down} := \True$, $\mathit{cross} := \False$\;
      \ForEach{$s \in S_?$\label{alg:OVI:Foreach}}{
        $v_\mathit{new} := \opt_{\mu \in T(s)} \sum_{\tuple{r, s'} \in \support{\mu}}{\mu(s') \cdot (r + v(s'))}$\label{alg:OVI:UpdateV}\;
        $u_\mathit{new} := \opt_{\mu \in T(s)} \sum_{\tuple{r, s'} \in \support{\mu}}{\mu(s') \cdot (r + u(s'))}$\label{alg:OVI:UpdateU}\;
        \lIf{$v_\mathit{new} > 0$}{%
          $\mathit{error} := \max\,\set{\mathit{error}, (v_\mathit{new} - v(s)) / v_\mathit{new} }$\label{alg:OVI:Error}
        }
        \lIf{$u_\mathit{new} < u(s)$}{$\mathit{up} := \False$\tcp*[f]{upper value decreased}}
        \lElseIf{$u_\mathit{new} > u(s)$}{$\mathit{down} := \False$\tcp*[f]{upper value increased}}
        \lIf{$u_\mathit{new} < v_\mathit{new}$}{$\mathit{cross} := \True$\tcp*[f]{upper value below lower}}
        $v(s) := v_\mathit{new}$, $u(s) := u_\mathit{new}$
      }
      \lIf{$\mathit{up} \vee \mathit{cross}$}{\Break\label{alg:OVI:NotUpper}\tcp*[f]{$u$ is definitely not an upper bound}}
      \ElseIf(\tcp*[f]{$u$ is an upper bound}){$\mathit{down} \wedge u(s_I) - v(s_I) \leq 2\epsilon \cdot v(s_I)$\label{alg:OVI:Upper}}{%
        \Return{$\frac{1}{2}(u(s_I) + v(s_I))$}\tcp*{and we have converged}
      }
    }
    $\mathit{error} := \frac{1}{2} \mathit{error}$\label{alg:OVI:DecrErr}\tcp*{decrease error for next iteration phase}
  }
}
\caption{Optimistic value iteration}
\label{alg:OVI}
\end{algorithm}

\noindent{}The resulting procedure in more detail is shown as \Cref{alg:OVI}.
Starting from the same initial vectors $v$ as for VI, we first perform standard Gauss-Seidel value iteration (in line~\ref{alg:OVI:CallVI}).
We refer to this as the \emph{iteration phase} of OVI.
After that, vector $v$ is an improved, and in practice usually very close, underapproximation of the actual probabilities or reward values.
We then ``guess'' an overapproximating vector $u$ of \emph{upper values} from the \emph{lower values} in $v$ by adding to $v$ the desired relative error, i.e.\ we multiply $v$ element-wise by $1 + \epsilon$ to obtain $u$ (line~\ref{alg:OVI:Guess}).
Then the \emph{verification phase} starts (in line~\ref{alg:OVI:Verif}):
we now perform value iteration on both the lower values $v$ and the upper values $u$ at the same time, keeping track of the direction in which the upper values move.
If, in some iteration, the upper values \emph{for all states moved down} (line~\ref{alg:OVI:Upper}), then we know by Park induction that the current $u$ is an inductive upper bound for the values of all states, see \Cref{eq:park}, and the true value must be in the interval $[v(s_I), u(s_I)]$.
If the interval is small enough \wrt $\epsilon$ (line~\ref{alg:OVI:Upper} checks a relative-width requirement of $2\epsilon$), then we can return its centre $v_I = \frac{1}{2}(u(s_I) + v(s_I))$ and be sure that the true value $v_\True = (\lfp \Phi)(s_I)$ is in $[v_I - \epsilon \cdot v_\True, v_I + \epsilon \cdot v_\True]$.
Otherwise, we remain in the verification phase---effectively performing interval iteration---until the two vectors $v$ and $u$ are sufficiently close.
If, on the other hand, the upper values for all states moved \emph{up}, or if we have $u(s) < v(s)$ for some state $s$, then the current $u$ is not an inductive upper bound. In line~\ref{alg:OVI:NotUpper}, we then cancel verification and go back to the iteration phase to further improve $v$ before trying again.

\paragraph{Optimisation.}
Recall that Park induction reads $\Phi(u) \preceq u$ implies $\lfp \Phi \preceq u$. 
Conversely---in case the fixed point of $\Phi$ is \emph{unique}---$u \preceq \Phi(u)$ implies that $u$ is a lower bound on $\lfp \Phi$.
In such situations of single fixed points, we can---as an optimisation---additionally replace $v$ by $u$ if all upper values have moved up at some point in the verification phase and continue with the iteration phase.

\paragraph{Heuristics.}
OVI is inherently a \emph{practical} approach that relies extensively on heuristics to gain an advantage over alternative methods such as II or SVI; it cannot be better on \emph{all} MDP.
Concretely, an implementation of OVI can choose
\begin{enumerate}
\item
a stopping criterion for the iteration phase,
\item
how to guess candidate upper values from the result of the iteration phase,~and
\item
how much to increase the ``precision'' requirement when going back from verification to iteration.
\end{enumerate}
\Cref{alg:OVI} shows the default choices made by our current implementation:
It (1.)~uses \Cref{alg:VI} and its standard relative-error stopping criterion for the iteration phase, but can be configured to use the absolute-error method instead.
We (2.)~guess upper values as shown in line~\ref{alg:OVI:Guess} if $\epsilon$ specifies a relative width; if an absolute width is required instead, then we simply add $\epsilon$ to all values in $v$.
In case of probabilities, we additionally replace values greater than $1$ by $1$ (not shown in \Cref{alg:OVI}).
Finally, when (3.)~going back to the iteration phase, we use half the error of the last iteration in the verification phase as the next value of the $\epsilon$ parameter of \texttt{GSVI} (as shown in line~\ref{alg:OVI:DecrErr}).
Reducing the error too much may cause more and potentially unnecessary iterations in \texttt{GSVI} (continuing to iterate although switching to the verification phase would already result in upper values sufficient for termination), while using too high a value may result in more verification phases (whose iterations are computationally more expensive than those of \texttt{GSVI}) being started before the values in $v$ are high enough.

\begin{example}
\label{ex:OVI}
We now call $\texttt{OVI}(M_e^0, \set{s_0, s_1, s_2}, \max, \set{ s_+ \mapsto 1 } \cup \set{ s \mapsto 0 \mid s \neq s_+ }, 0.05)$.
\Cref{tab:ExampleOVI} shows the values in $v$ and $u$ during this run, using an absolute-width requirement of $\epsilon = 0.05$ and the absolute-error stopping criterion in \texttt{GSVI}.
The first iteration phase lasts from $i = 0$ to~$4$.
At this point, $u$ is initialised with the values shown in italics.
The first verification phase needs only one iteration to realise that $u$ is actually a lower bound (to a fixed point which is not the least fixed point, due to the uneliminated end component).
We then resume \texttt{GSVI} from $i = 6$.
The error in \texttt{GSVI} is again below $\epsilon_\mathit{VI}$, which had been reduced to $0.008$, during iteration $i = 9$.
We thus start another verification phase, which immediately (in one iteration) finds the newly guessed vector $u$ to be an upper bound, and the difference between $u(s_0)$ and $v(s_0)$ to be small enough.
\end{example}

\paragraph{Termination.}
We first consider situations where the Bellman operator has a single fixed point (which is always achievable by precomputations and transformations as described in \Cref{sec:Preliminaries}).
At some point, all values in $v$ will be close enough to the true values so that the guessing phase picks a \emph{valid} upper bound $u \succeq \lfp \Phi$.
However, this bound need not be \emph{inductive}, i.e.\ $\Phi(u) \not\preceq u$.
Moreover, even in the case where value iteration on this upper bound does converge towards the least fixed point\footnote{This could even be guaranteed by modifying VI to enforce monotonicity as in~\cite{BKLPW17}.}, i.e.\ if $\lim_{n \to \NN} \Phi^n(u) = \lfp \Phi$, the improved upper bound may never become inductive, i.e.\ there may not exist an $n$ such that $\Phi^{n+1}(u) \preceq \Phi^n(u)$.
In this scenario, OVI will remain unaware that it has in fact picked a valid upper bound~$u$ because it is unable to prove this fact by Park induction.
Thus, if the guessing heuristics of OVI continually picks such an unfavourable $u$, then OVI will not terminate.
\emph{In practice}, however, we have not yet encountered such a situation outside of constructed examples with \emph{constructed vectors} $u$ that OVI with our implemented guessing heuristics could not have chosen.

%

If we apply OVI in a situation with multiple fixed points (\eg by skipping the precomputations, or by not computing and eliminating end components\footnote{We must however ensure that at least the \emph{least} fixed point corresponds to the true~values, \ie we must eliminate end components for $\mathrm{E}_\min$ properties---but only for those.}), then we can additionally get nontermination due to the guessed upper values \emph{always} moving \emph{up} towards a higher fixed point, resulting in infinitely many validation phases being cancelled.
The situation where they move \emph{down} in verification phase iteration $i$, but another fixed point exists between $u$ and the true values, is only problematic with our guessing heuristics if additionally values moved up in iterations $j < i$ such that the difference between $v(s_I)$ and $u(s_I)$ forever remains higher than the required width.
Again, we have not encountered either situation on practical examples yet.
To mitigate (but not eliminate) the second case in models yet unknown to us, as well as the case of never reaching an inductive upper bound described in the previous paragraph, our implementation additionally cancels verification when the current verification phase took more than ten times as many iterations as the previous iteration phase.

In summary, OVI is a semi-algorithm: it need not terminate.
On all MDP that we have tested, however, it does terminate.
This, together with the importance of heuristics, again underlines the practical nature of OVI.

\section{Experimental Evaluation}
\label{sec:Experiments}

We have implemented interval iteration (II) (using the ``variant 2'' approach of~\cite{BKLPW17} to compute initial overapproximations for expected rewards), sound value iteration (SVI), and now optimistic value iteration (OVI) precisely as described in the previous section, in the \mcsta model checker of the \toolset~\cite{HH14}, which is publicly available at \href{http://www.modestchecker.net/}{modestchecker.net}.
It is cross-platform, implemented in C\#, and built around the \modest~\cite{HHHK13} high-level modelling language.
Via support for the \jani format~\cite{BDHHJT17}, the toolset can exchange models with other tools like \tool{Storm}~\cite{DJK017} and \tool{Epmc}~\cite{HLSTZ14}.
\mcsta is the toolset's explicit-state probabilistic model checker.
Its performance is competitive with \tool{Storm} and \tool{Prism}~\cite{HHHKKKPQRS19}.

\begin{figure}[t]
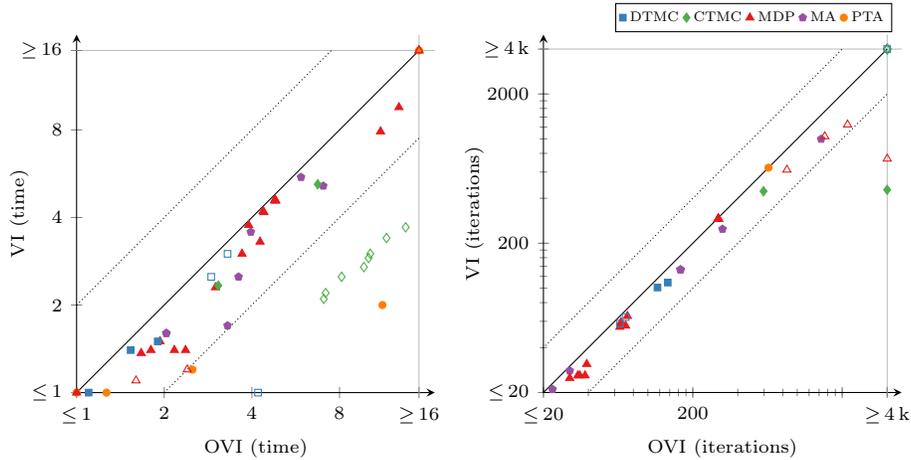

\centering
\scatterplotpvi{results-p.csv}{mcsta.ovi.std}{OVI (time)}{mcsta.vi.std}{VI (time)}{0.52\textwidth}{false}%
\scatterplotiters{results-p.csv}{mcsta.ovi.std}{OVI (iterations)}{mcsta.vi.std}{VI (iterations)}{0.52\textwidth}{true}%
\caption{OVI runtime and iteration count compared to VI (probabilistic reachability)}
\label{fig:PlotsPVI}
\end{figure}

\begin{figure}[t]
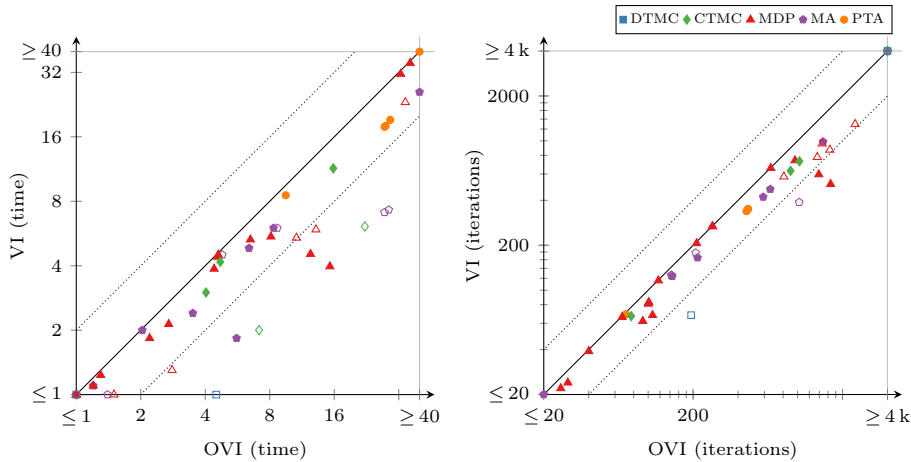

\centering
\scatterplotevi{results-e.csv}{mcsta.ovi-vi.std}{OVI (time)}{mcsta.vi.std}{VI (time)}{0.52\textwidth}{false}%
\scatterplotiters{results-e.csv}{mcsta.ovi.std}{OVI (iterations)}{mcsta.vi.std}{VI (iterations)}{0.52\textwidth}{true}%
\caption{OVI runtime and number of iterations compared to VI (expected rewards)}
\label{fig:PlotsEVI}
\end{figure}

In the following, we report on our experimental evaluation of OVI using \mbox{\mcsta} on all applicable models of the Quantitative Verification Benchmark Set (QVBS)~\cite{HKPQR19}.
All models in the QVBS are available in \jani and can thus be used by \mcsta.
Most of them are parameterised, and come with multiple properties of different types.
Aside from MDP models, the QVBS also includes DTMCs (which are a special case of MDP), continuous-time Markov chains (CTMC, for which the analysis of unbounded properties reduces to checking the embedded DTMC), Markov automata (MA~\cite{EHZ10}, on which the embedded MDP suffices for unbounded properties), and probabilistic timed automata (PTA~\cite{KNSS02}, some of which can be converted into MDP via the digital clocks semantics~\cite{KNPS06}).
We use all of these model types.
The QVBS thus gives rise to a large number of benchmark \emph{instances}:
combinations of a model, a parameter valuation, and a property to check.
For every model, we chose a representative set of instances, aiming to cover all its unbounded probabilistic reachability and expected-reward properties as well as one or two suitable parameter valuations.
We only excluded
\begin{itemize}
\item 
models with multiple initial states (which \mcsta does not yet support),
\item 
probabilistic reachability properties for which the result is $0$ or $1$ (since they can easily be solved by the graph-based precomputations),
\item
the \textit{oscillators} model due to its very large model files,
\item
model-property combinations for which we found no parameter valuation s.t.
\begin{itemize}
\item[--]
VI, II, SVI, or OVI took more than 1 second (since lower runtimes do not allow reliable comparisons) and
\item[--]
the entire model checking process (including state space exploration) did not run out of memory or exceed a 2-minute timeout.
\end{itemize} 
\end{itemize}
As a result, we considered 47 instances with probabilistic reachability and 47 instances with expected-reward properties.
For many of them, ``reference results'' are available; in those cases, we also checked that the result delivered by the respective method is correct up to the requested error width.

We ran all experiments on an Intel Core i7-4790 workstation ($3.6$-$4.0\sunit{GHz}$) with 8\sunit{GB} of memory and 64-bit Ubuntu Linux 18.04, using version 3.1 of the \toolset.
We request a relative half-width of $\epsilon = 10^{-6}$ for the result probability or reward value, and configure OVI to use the relative-error criterion with $\epsilon_\mathit{VI} = 10^{-6}$ in the iteration phase.
We report the average over three runs for every instance.
Due to the number of instances, we show the results of our experiments as scatter plots like in \Cref{fig:PlotsPVI}.
Each such plot compares two methods in terms of runtime or number of iterations.
Every point $\tuple{x, y}$ corresponds to an instance and indicates that the method noted on the x-axis took $x$ seconds or iterations to solve this instance while the method noted on the y-axis took $y$ seconds or iterations.
Thus points above the solid diagonal line correspond to instances where the x-axis method was faster (or needed fewer iterations); points above (below) the upper (lower) dotted diagonal line are where the x-axis method took less than half (more than twice) as long or as many iterations.

\paragraph{Comparison with VI.}
All methods except VI delivered correct results, \ie within $\pm\,\epsilon \cdot r$ where a reference result $r$ is available.
VI offers low runtime at the cost of occasional incorrect results, and in general the absence of any guarantee about the result.
We thus compare with VI separately to judge the overhead caused by performing additional verification, and possibly iteration, phases.
\Cref{fig:PlotsPVI,fig:PlotsEVI} show the results.
The unfilled shapes indicate instances where a reference result is available and VI produced an incorrect result.
In terms of runtime, we see that OVI does not often take more than twice as long as VI, and in most cases requires less than $50\,\%$ extra time.
On many of the instances where OVI incurs a significant overhead, VI produces an incorrect result, indicating that they are ``hard'' instances for value iteration.
The unfilled CTMCs where OVI takes much longer to compute probabilities are all instances of the \textit{embedded} model; the DTMC on the x-axis is \textit{haddad-monmege}, an adversarial model built to highlight the convergence problem of VI in~\cite{HM14}.
The problematic cases for expected rewards include the two instances of the \textit{ftwc} MA model, the two expected-reward instances of the \textit{embedded} CTMC, and again \textit{haddad-monmege}.
In terms of iterations, the overhead of OVI is even less than in runtime.
When inspecting the output of \mcsta, we found that OVI usually requires few very short verification phases.

\begin{figure}[t]
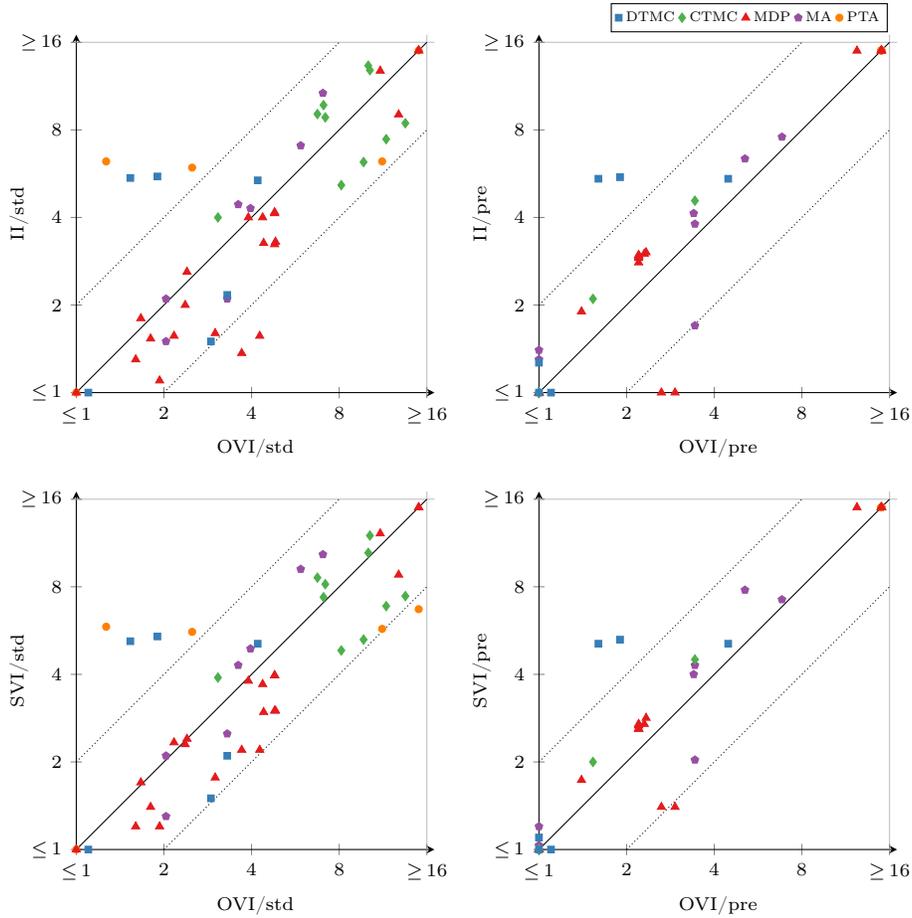

\centering
\scatterplotp{results-p.csv}{mcsta.ovi.std}{OVI/std}{mcsta.ii.std}{II/std}{0.52\textwidth}{false}%
\scatterplotp{results-p.csv}{mcsta.ovi.pre}{OVI/pre}{mcsta.ii.pre}{II/pre}{0.52\textwidth}{true}\\[5pt]
\scatterplotp{results-p.csv}{mcsta.ovi.std}{OVI/std}{mcsta.svi.std}{SVI/std}{0.52\textwidth}{false}%
\scatterplotp{results-p.csv}{mcsta.ovi.pre}{OVI/pre}{mcsta.svi.pre}{SVI/pre}{0.52\textwidth}{false}%
\caption{OVI runtime compared to II and SVI (probabilities)}
\label{fig:PlotsP}
\end{figure}

\begin{figure}[t]
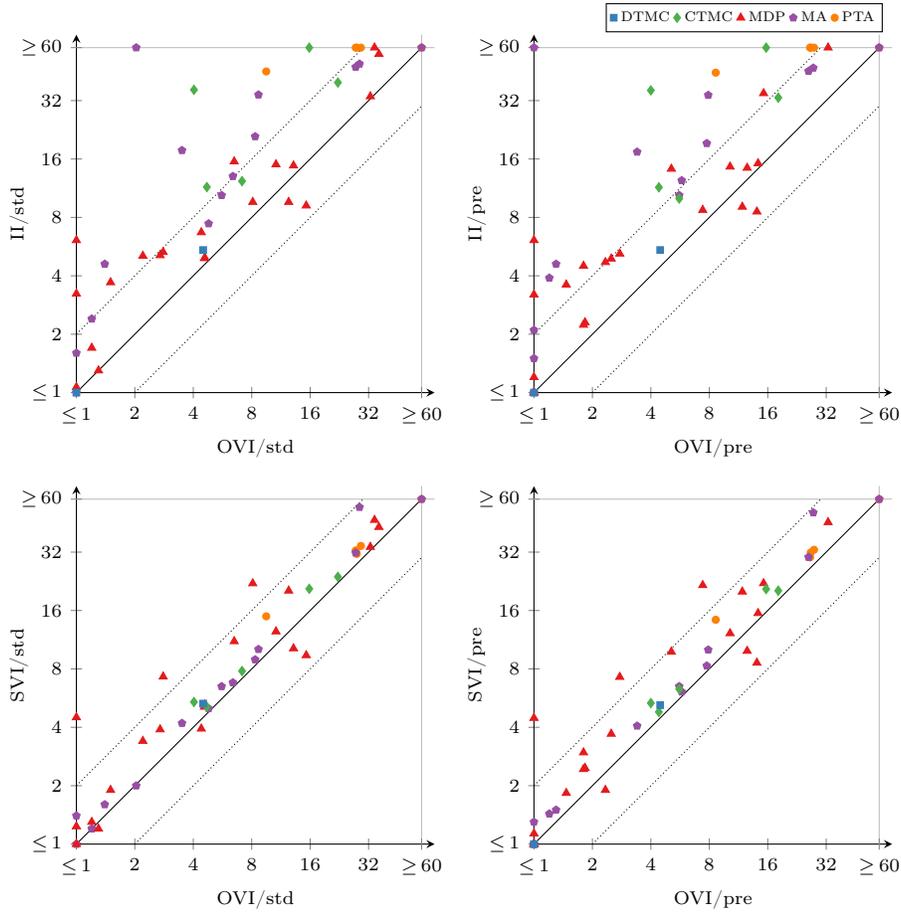

\centering
\scatterplote{results-e.csv}{mcsta.ovi.std}{OVI/std}{mcsta.ii.std}{II/std}{0.52\textwidth}{false}%
\scatterplote{results-e.csv}{mcsta.ovi.pre}{OVI/pre}{mcsta.ii.pre}{II/pre}{0.52\textwidth}{true}\\[5pt]
\scatterplote{results-e.csv}{mcsta.ovi.std}{OVI/std}{mcsta.svi.std}{SVI/std}{0.52\textwidth}{false}%
\scatterplote{results-e.csv}{mcsta.ovi.pre}{OVI/pre}{mcsta.svi.pre}{SVI/pre}{0.52\textwidth}{false}%
\caption{OVI runtime compared to II and SVI (expected rewards)}
\label{fig:PlotsE}
\end{figure}

\paragraph{Comparison with II and SVI.}
We compare the runtime of OVI with the runtime of II and that of SVI separately for reachability probabilities (shown in \Cref{fig:PlotsP}) and expected rewards (shown in \Cref{fig:PlotsE}).
OVI has the same requirements on precomputations as VI (\ie only end component elimination is needed only for $\mathrm{E}_\min$ properties), while II and SVI require the use of precomputations and of end component elimination (for $\mathrm{P}_{\!\!\max}$ properties) as discussed in \Cref{sec:Preliminaries}.
The precomputations and end component elimination need extra runtime (which turned out to be negligible in some cases but significant enough to cause a timeout in others) prior to the numeric iterations.
However, doing the precomputations can reduce the size of the set $S_?$, and end component elimination can reduce the size of the MDP itself.
Both can thus reduce the runtime needed for the numeric iterations.
For the overall runtime, we found that none of these effects dominates the other over all models.
Thus sometimes it may be better to perform only the required precomputations and transformations, while on other models performing all applicable ones may lead to lower total runtime.
We thus compare OVI, II, and SVI in two scenarios:
once in the default (``std'') setting of \mcsta that uses only required precomputations and transformations (where we report the total runtime for precomputations, transformations, and numeric iterations), and once with all of them enabled (``pre'', where we report only the runtime for numeric iterations, plus the computation of initial upper bounds in case of~II).

For probabilistic reachability, we see in \Cref{fig:PlotsP} that there is no clear winner among the three methods in the ``std'' setting.
We found that, for the QVBS models, value iteration to compute probabilities is usually fast, and the overall model checking time is dominated by the time needed for state space exploration.
We were unable to scale up several models to require more than $1\sunit{s}$ for value iteration without running out of memory due to the state space exploding.
Similarly, the precomputations and transformation take relatively long enough to significantly influence the outcome.
The ``pre'' setting, in which all three algorithms operate on exactly the same input \wrt to MDP $M$ and set $S_?$, however, shows a clearer picture:
OVI is consistently faster than both II and SVI, with only 6 instances where it takes longer (which are the single instances of the \textit{stream} and \textit{nand} models as well as two instances each of \textit{csma} and \emph{zeroconf}).

Expected-reward properties were more challenging for all three methods (as well as for VI, which produced more errors here than for probabilities), and the precomputations and transformations have less impact on runtime.
The plots in \Cref{fig:PlotsE} paint a very clear picture of OVI being significantly faster for expected rewards than II (which suffers from the need to precompute initial upper bounds that then turn out to be rather conservative), and faster (though by a lesser margin) than SVI.
The outliers are the single instances of \textit{coupons} and \textit{polling-system}, one instance each of \textit{csma} and \textit{firewire}, and two instances of \textit{wlan}.

\section{Conclusion}
\label{sec:Conclusion}

We have presented \emph{optimistic value iteration} (OVI), a new approach to making non-exact probabilistic model checking via iterative numeric algorithms sound in the sense of delivering results within a prescribed interval around the true value (modulo floating-point and implementation errors).
Compared to the existing approaches of interval (II) and sound value iteration (SVI), OVI is \emph{theoretically} weaker since it cannot guarantee termination.
However, it is deeply \emph{practical}:
\begin{itemize}
\item 
It terminates on ``regular'' models, including on all applicable models and properties of the Quantitative Verification Benchmark Set.
\item
It relies on a combination of heuristics that can be arbitrarily modified and tuned, but that crucially determine its effectiveness and efficiency. 
\item
It is faster than II and SVI when computing probabilities on a ``level playing field'' (\ie modulo precomputations and transformations), and it is unconditionally faster than either of the two when computing expected rewards.
\item
It is very simple to add to any tool that already implements value iteration.
\end{itemize}
In summary, there is no more excuse for a probabilistic model checker (several of which still default to unsound VI due to the effort required to implement II or SVI) not to (try to) produce sound results now (via OVI).

\paragraph{Future work.}
We have so far implemented OVI (in \mcsta) with one set of heuristics as described in this paper.
While they turned out to work very well, making OVI faster than all current alternatives, we see ample room for improvement especially in devising better methods to guess the initial upper bounds for the verification phase, and in tuning how $\epsilon_\mathit{VI}$ is adjusted when going back to the iteration phase.
We also plan to run more extensive experiments, in particular comparing OVI across different absolute and relative-width requirements, and with initial values for $\epsilon_\mathit{VI}$ that differ from the specified half-width~$\epsilon$.

\paragraph{Acknowledgments.}
The authors thank Tim Quatmann (RWTH Aachen) for fruitful discussions when the idea of OVI initially came up in late 2018, and for his help in implementing and optimising the SVI implementation in \mcsta.

\bibliography{paper}
\bibliographystyle{splncs04}

\end{document}